\documentclass[a4paper]{article}
\usepackage{ASVspoof}
\usepackage{epsfig,amssymb,amsmath}
\usepackage{hyperref}
\usepackage{xcolor}
\usepackage{siunitx}
\usepackage{multirow}
\usepackage{booktabs}
\usepackage{pifont}% http://ctan.org/pkg/pifont
\newcommand{\cmark}{\ding{51}}%
\newcommand{\xmark}{\ding{55}}%
\usepackage{makecell}       % for multi-line table cells
\usepackage{xurl}
\newcolumntype{x}[1]{>{\centering\arraybackslash}p{1cm}}
\ninept

\title{Learn from Real: Reality Defender's Submission to ASVspoof5 Challenge}

\makeatletter
\def\name#1{\gdef\@name{#1\\}}
\makeatother
\name{{\em Yi Zhu\textsuperscript{1,2*\thanks{* \textdagger This work was done during internship at Reality Defender Inc.}}, Chirag Goel\textsuperscript{1,3\textdagger}, Surya Koppisetti\textsuperscript{1}, Trang Tran\textsuperscript{1}, Ankur Kumar\textsuperscript{1}, Gaurav Bharaj\textsuperscript{1}}}
\address{
$^1$Reality Defender, New York, USA  \\
$^2$Institut National de la Recherche Scientifique, Montréal, Canada\\
$^3$Université de Montréal, Montréal, Canada
}
\begin{document}
\maketitle

\begin{abstract}
Audio deepfake detection is crucial to combat the malicious use of AI-synthesized speech. Among many efforts undertaken by the community, the ASVspoof challenge has become one of the benchmarks to evaluate the generalizability and robustness of detection models. In this paper, we present Reality Defender's submission to the ASVspoof5 challenge, highlighting a novel pretraining strategy which significantly improves generalizability while maintaining low computational cost during training. Our system SLIM learns the style-linguistics dependency embeddings from various types of bonafide speech using self-supervised contrastive learning. The learned embeddings help to discriminate spoof from bonafide speech by focusing on the relationship between the style and linguistics aspects. We evaluated our system on ASVspoof5, ASV2019, and In-the-wild. Our submission achieved minDCF of 0.1499 and EER of 5.5\% on ASVspoof5 Track 1, and EER of 7.4\% and 10.8\% on ASV2019 and In-the-wild respectively. %Furthermore, we compare the improvements brought by different model design choices and discuss the potential causes of misclassifications.
\end{abstract}

\section{Introduction}
% Definition of audio deepfakes, motivations for deepfake detection.
The increasing interest in speech generative models has resulted in rapidly emerging text-to-speech (TTS) and voice conversion (VC) tools. With many tools being publicly available~\cite{almutairi2022review, masood2023deepfakes}, nowadays a person's voice can be cloned from a only few seconds of a speech recording. During recent years, there have been many cases of misusing such techniques, e.g., impersonation of celebrity's voice~\cite{news1}, hacking bank accounts using cloned voice over telephone line~\cite{news2}, evidence forgery at court~\cite{news3}, just to name a few.

% Summary of past ASVspoof challenges and intro of ASVspoof 5 challenge tracks.
To combat the spread of these maliciously generated speech (i.e., speech deepfakes), numerous efforts have been undertaken by the research and industry communities, including the ASVspoof series of challenges~\cite{todisco2019asvspoof, liu2023asvspoof}. The main objective of ASVspoof challenges is to encourage the innovation of speech deepfake detection tools that can generalize to unseen attacks and maintain robustness under various conditions, such as transmission and compression codecs. In Track 1 of the most recent ASVSpoof5 challenge (referred to henceforth as ASV5), the  goal is to build a standalone bonafide vs. spoof deepfake detection system. 

%The most recent ASVspoof5 challenge (hereinafter abbreviated as ASV5) entails two challenge tracks, namely a stand-alone speech deepfake (bonafide v.s. spoof) detection task (track 1) and a spoofing-robust automatic speaker verification task (track 2). Since our team participated exclusively in track 1 open condition, the following text will concentrate solely on the corresponding task. 

% our submission
Existing state-of-the-art (SOTA) systems on deepfake detection typically adopt one or more self-supervised learning (SSL) speech encoders as \emph{feature extraction} frontend, and append various \emph{downstream} classifiers as backend~\cite{almutairi2022review, masood2023deepfakes}. The majority of the innovations focus solely on the \emph{downstream} supervised training stage, including full finetuning large SSL encoders~\cite{9747768, tak2022automatic}, designing classifiers with more discriminative power~\cite{jung2022aasist, kang2024experimental}, increasing the variety of spoof training data~\cite{wang2023spoofed, wang2024can, tak2022rawboost}, and model ensembling~\cite{yang2024robust}. While demonstrating improvement on some datasets, the training cost of these methods can be high, especially considering that the variety of speech generative models has drastically increased over time. Taking ASV5 as an example, a single epoch of training a model with a frozen Wav2vec-Large frontend and a simple classification backend (5 million parameters) can take around \SI{4}{\hour} on one A100 GPU, whereas the same setup took less than \SI{30}{\minute} for previous ASVspoof challenges. Worse, methods that finetune models end-to-end would require several rounds of hyperparameter search, and still might underperform models using a frozen backbone, as noted from the trends in ASV5 \cite{Wang2024_ASVspoof5}. 

In this paper, we present a summary of Reality Defender's submission to the {\it eval} phase under  Track 1 of ASV5. 
We used a novel pretraining framework SLIM, originally proposed in our recent work~\cite{zhu2024slim}, which exploits the mismatch between style and linguistics content in deepfake data to detect them. SLIM involves two stages of training: the first stage adopts self-supervised contrastive learning on real data to learn the style-linguistics dependency embeddings; the second stage leverages the embeddings learned from the first stage and trains a classifier in a supervised manner to separate real from deepfake speech. With a low training cost (7 million trainable parameters; less than \SI{15}{\hour} training time including pretraining,\footnote[1]{Details of compute can be found in Appendix.~\ref{appendix:1}}) SLIM achieved competitive results on ASV5. Our test results also show that the model generalizes well to out-of-domain datasets.

The rest of the paper is organized as follows. In Section 2, we introduce the employed speech deepfake datasets for model evaluation, including ASV5 data and two other datasets. Section 3 provides a description of our submitted system, including the pretraining strategies and downstream finetuning details. Section 4 discusses the results achieved on the three test datasets. Section 5 presents the conclusions. 

\section{Datasets and evaluation metrics}
Following the challenge rules, we used the official ASV5 track 1 data for training and evaluated our system by submitting scores for the {\em eval} dataset on the CodaLab platform.\footnote[2]{\url{https://codalab.lisn.upsaclay.fr/competitions/19380}} We used ASV2019 Logical Access (LA)~\cite{todisco2019asvspoof} and In-the-wild (ITW)~\cite{muller2022does} datasets as two out-of-domain (OOD) corpora for offline evaluation of the model generalizability. Table~\ref{tab:dataset} summarizes the statistics of employed datasets and data partitions.
For details on the ASVSpoof5 challenge, and the permissible data constraints within each track, please see the papers~\cite{Wang2024_ASVspoof5, todisco2019asvspoof, muller2022does}.
% Prefix `ASVspoof' is omitted for row 1-5 and only the year of challenge is displayed. 
\setlength{\tabcolsep}{1.8pt}
\begin{table}[hbpt]
\small
\caption{\label{dataset} {\it Summary of our train, validation, and test datasets. ``Dur.'' denotes average duration (in seconds) of samples in each set. Unknown details on ASV5 {\em prog} and {\em eval} are marked as NA. \textsuperscript{*}OOD datasets that are not part of ASV5.} }
\vspace{2mm}
\centering
% \begin{tabular}{x{0.8cm}x{3cm}x{0.8cm}x{0.8cm}x{0.8cm}x{0.8cm}}
\begin{tabular}{cccccc}
\toprule
Partition & Source & Dur.\ (s) & \#Bonafide & \#Spoof & \#Attacks \\
\midrule
Train & ASV5 {\em train} & 11.9 & 18797 & 163560 & 8 \\
Valid & ASV5 {\em dev} & 7.1 & 31334 & 109616 & 8 \\
\midrule
Test & ASV5 {\em prog} & 7.1& NA & NA & NA\\
Test & ASV5 {\em eval} & 7.1& 138688 & 542086 & 16\\
Test\textsuperscript{*} & ASV19 {\em eval} & 3.1 & 7355 & 63882 & 13 \\
Test\textsuperscript{*} & ITW & 4.3 & 19963 & 11816 & NA\\
\bottomrule
\end{tabular}
\label{tab:dataset}
\end{table}

\subsection{Datasets}
\textbf{ASV5 track 1}: Four sets of data were released under track 1: training (\emph{train}), development (\emph{dev}), progress (\emph{prog}), and evaluation (\emph{eval}). The {\em prog} set, a subset of the {\em eval} set, was released during the progress phase for participants to determine the best model candidates for the final submission. As outlined in the post-challenge review from the organizers \cite{Wang2024_ASVspoof5}, utterances in the {\em eval} set exhibit varied signal quality due to the application of speech coding, audio compression, bandlimiting, or other processing algorithms. The ASV5 {\em eval} set was our primary test set.\\
\textbf{ASVspoof2019 LA (ASV2019)}: As our second test set, we use the ASV2019 LA {\em eval} set, which spans 13 types of spoofing attacks. Although the setup of ASV2019 LA was similar to that of ASV5, the average recording duration of ASV2019 {\em eval} (\SI{3.1}{\second}) is markedly shorter than that of ASV5 {\em train} (\SI{11.9}{\second}). \\
\textbf{In-the-wild (ITW)}: Our third test set, ITW, contains audio clips from English-speaking celebrities and politicians. When compared to ASV2019, the ITW set features more realistic and spontaneous speech samples, with more complicated acoustic environment (e.g., noise from crowds, reverberated speech). The bonafide and spoof classes have a better balance in ITW.

\subsection{Metrics}
The minimum detection cost function (minDCF) was used as the main metric together with equal error rate (EER). Log-likelihood ratio (LLR) was used a complementary metric to measure the confidence level of the model. For evaluation and comparison on ASV2019 and ITW, we use the EER metric.

\section{System description}
This section describes the architecture and the two-stage training framework of SLIM, which in general follows the same methodology as proposed in \cite{zhu2024slim}. Adjustments have been made to our submitted system to ensure SLIM's adherence to the challenge rules. These adjustments are described in \ref{ssec:sscl}.%next.

\subsection{Overview}

\begin{figure*}
\centering
\includegraphics[width=0.85\linewidth]{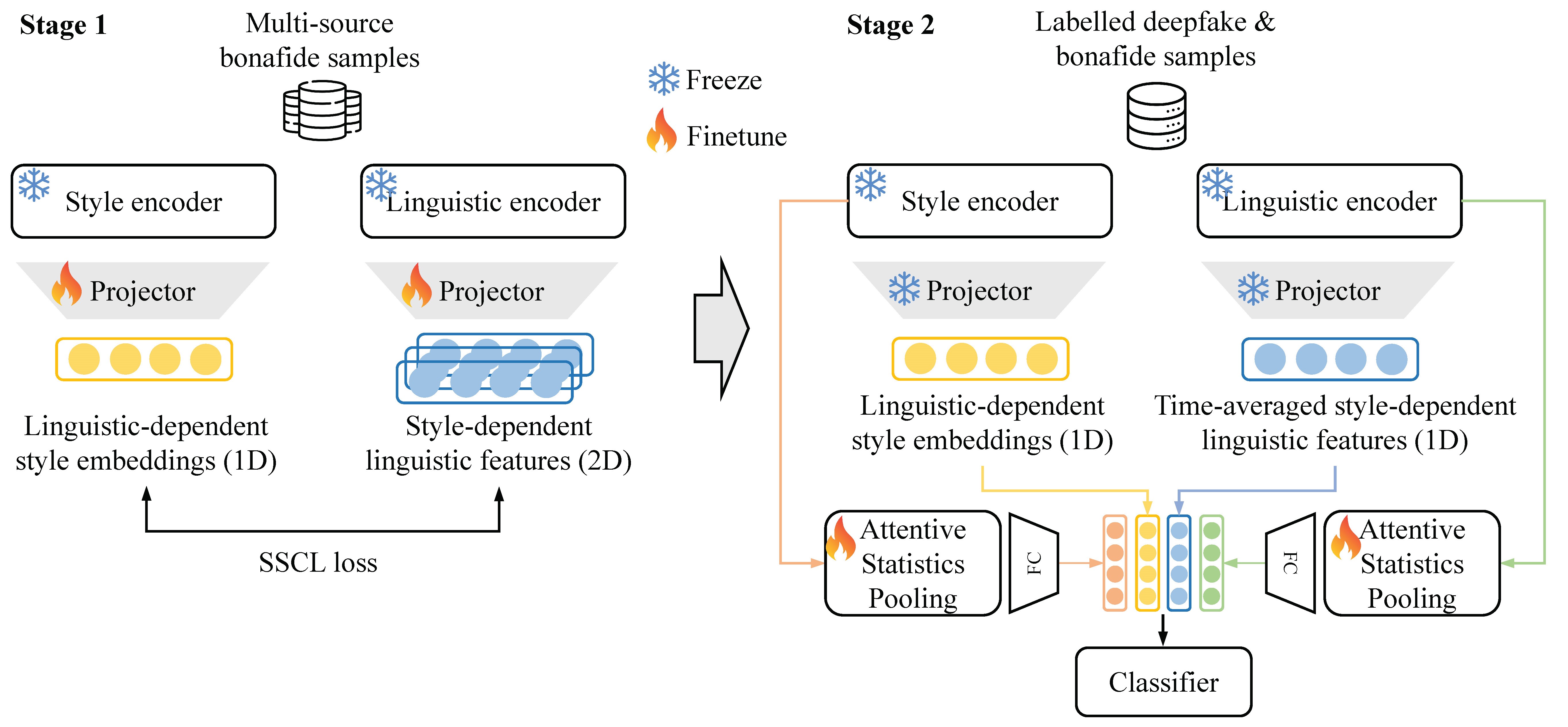}
\caption{{\it Two-stage training framework of SLIM. Stage 1 extracts style and linguistics representations from frozen SSL encoders, projects them into a lower-dimensional space, and aims to minimize the distance between the projected representations as well as the intra-subspace redundancy. The Stage 1 embeddings, style embeddings (output from style encoder) and linguistics embeddings (output from linguistics encoder) are concatenated in Stage 2 to learn a classifier via supervised training. Architecture of the projector network and classifier can be found in Appendix.~\ref{appendix:2}}}
\label{fig:system}
\end{figure*}

The general training framework is depicted in Figure.~\ref{fig:system}. SLIM follows a two-stage training process, where the first stage adopts self-supervised contrastive learning (SSCL) to model style-linguistics dependency from various types of bonafide speech, while the second stage learns to utilize the style-linguistics relationship to further discriminate spoof from bonafide via standard supervised training. 

The objective of the first stage is to learn embeddings that capture the dependencies between the style and linguistics aspects in real speech. {\em Style} is assumed to encompass short- and long-term paralinguistic attributes, including speaker identity, emotion, accent, and health state~\cite{schuller2013paralinguistics}. {\em Linguistics} refers to the verbal content of speech~\cite{kretzschmar2009linguistics}. While the two are often considered as independent speech aspects during speech modelling, studies have demonstrated a specific dependency between these two subspaces, such as the connection between emotional states and word choices~\cite{lindquist2015role}, the relationship between prosody and language comprehension~\cite{cutler1997prosody}, and the influence of age on sentence coherence~\cite{pereira2019age}. While mainstream SSL models, such as Wav2vec~\cite{baevski2020wav2vec}, WavLM~\cite{chen2022wavlm}, Data2vec~\cite{baevski2022data2vec}, have been shown to encode the style and linguistics information in different transformer layers~\cite{cho2023evidence, pasad2023comparative, saliba2024layer}, the cross-subspace dependency is not well modelled with these SSL representations, which can be crucial for discriminating spoof from bonafide~\cite{zhu2024slim}. Therefore, we propose to bridge this gap by learning a set of style-linguistics dependency embeddings at Stage 1, and fuse these learned features with the original SSL representations to enhance their discriminative power.

\subsection{Self-supervised contrastive learning}
\label{ssec:sscl}
The objective of the first stage is to learn pairs of style and linguistics features that are expected to be highly correlated for real speech and minimally correlated for deepfakes. Since only bonafide speech samples are required at this stage, we employed subsets from CommonVoice~\cite{ardila2020common} and RAVDESS~\cite{livingstone2018ryerson} datasets as the Stage 1 training data. The former contains a large number of speakers, and the latter has utterances in different emotional states, hence their combination forms a training set with diverse style traits.

We first extract style and linguistics representations using pretrained SSL backbones. In \cite{zhu2024slim}, SLIM relied on layer 0-11 and layer 12-22 from Wav2vec2-XLSR finetuned for speech emotion recognition and speech recognition as style and linguistics representations, respectively. The choice of layers was based on existing findings showing that early layers in SSL backbones are highly correlated with speaker attributes and later layers with verbal content~\cite{saliba2024layer, cho2023evidence, pasad2023comparative}. Since the XLSR backbones do not comply with the ASV5 rules, we experimented with representations using other challenge-approved SSL speech encoders, such as Wav2vec2-Large, Wav2vec2-Base, WavLM-Base, and Data2vec-Base. Our experiments showed best results using layers 0-7 and layers 8-11 from WavLM-Base, so we proceeded with WavLM-Base backbones. 

Both style and linguistics representations are three-dimensional tensors $\in \mathbb{R}^{L\times F \times T}$, where $L$ denotes the number of transformer layers, $F$ the feature size, and $T$ the number of time steps. The two subspace representations are then sent into projector networks ($\mathcal{P}$), which average the transformer layer outputs and reduce the feature size from 768 to 256 (see Appendix~\ref{appendix:2} for the architecture of the projector network). The output from the projector networks are regarded as dependency features: $\mathbf{S}_{f,t} = \mathcal{P}(\mathbf{X}_S)$ for style and $\mathbf{L}_{f,t} = \mathcal{P}(\mathbf{X}_L)$ for linguistics, and their temporally averaged versions are denoted $\bar{\mathbf{S}}_{f}$ and $\bar{\mathbf{L}}_{f}$. These dependency features are learned by minimizing the self-supervised contrastive loss $\mathcal{L}_{SSC}$, which comprises two terms: 
\begin{equation}
    \mathcal{L}_{SSC} = \mathcal{L}_{D} + \mathcal{\lambda L}_{R}
\end{equation}
$\mathcal{L}_{D}$ represents the distance between the projected style and linguistics features, $\mathcal{L}_{R}$ represents the self-redundancy of the learned features, and $\lambda \in [0,1]$ is a hyperparameter that weighs the two loss terms, defined as follows: 

\begin{equation}
    \mathcal{L}_{D} = \frac{1}{T} \sum_{t=0}^{T}\|\mathbf{S}_{f,t}-\mathbf{L}_{f,t}\|^{2}_{\mathbf{F}},
\end{equation}

\begin{equation}
    \mathcal{L}_{R}  = \|\mathbf{\bar{S}}_{f}\mathbf{\bar{S}}^\intercal_{f}-\mathbb{I}\|^{2}_{\mathbf{F}} + \|\mathbf{\bar{L}}_{f}\mathbf{\bar{L}}^\intercal_{f}-\mathbb{I}\|^{2}_{\mathbf{F}}
\end{equation}
where $T$ is the number of time steps; and $\|(.)\|^{2}_{\mathbf{F}}$ is the Frobenius norm. The $\mathcal{L}_{D}$ term reduces distance between the projected style and linguistic embeddings, the $\mathcal{L}_{R}$ term reduces redundancy within the (temporally averaged) style and linguistic features by pushing off-diagonal elements to zero. The PyTorch-style implementation of the SSC loss can be found in \cite{zhu2024slim}.

\subsection{Supervised downstream finetuning}
The second stage follows a standard supervised training approach based on features learned from Stage 1. Since the dependency features are designed to capture solely the style-linguistics mismatch, deepfake-related artifacts may be neglected. Hence, we complement them with the raw SSL embeddings extracted from the pretrained WavLM-Base backbone to increase the discriminative power. As shown in Figure~\ref{fig:system}, SSL embeddings are first passed through an attentive statistics pooling (ASP) layer, followed by a fully-connected layer to align with the dimension of the dependency features (256-dim). All features are then concatenated and fed into the downstream classifier to obtain a final output. Details of the classifier can be found in Appendix.~\ref{appendix:2}

% % \setlength{\tabcolsep}{3pt}
% \begin{table}[]
% \caption{\label{encoders} {\it Experimented base speech SSL encoders for style and linguistics representations. Last column indicates whether the representation is adopted in the final submission system.}}
% \vspace{2mm}
% \centering
% \begin{tabular}{cccc}
% \toprule
% Backbone & \#Layer & Subspace & Used\\
% \midrule
% Wav2vec-Large & 0-11 & Style & \xmark \\
% Wav2vec-Base & 0-7 & Style & \xmark\\
% WavLM-Base & 0-7 & Style & \cmark\\
% Data2vec-Base & 0-7 & Style & \xmark\\
% Wav2vec-Large & 12-22 & Linguistics & \xmark\\
% Wav2vec-Base & 8-11 & Linguistics & \xmark\\
% WavLM-Base & 8-11 & Linguistics & \cmark\\
% Data2vec-Base & 8-11 & Linguistics & \xmark\\
% \bottomrule
% \end{tabular}
% \label{tab:encoders}
% \end{table}

\subsection{Training details}
\emph{Preprocessing}: All speech recordings were loaded with the \textsc{Torchaudio} library~\cite{yang2022torchaudio}, then resampled to \SI{16}{kHz} with amplitude normalized between -1 and 1. This step was applied to all samples in train, validation, and test sets. For training efficiency and memory constraints, we truncate waveforms that are longer than \SI{10}{\second}. During validation and inference, all samples were kept at their original lengths.\\
\emph{Data augmentation}: No data augmentation was applied during Stage 1 SSCL training. RawBoost~\cite{tak2022rawboost} augmented samples were concatenated with original samples during Stage 2 supervised training to combat potential loss brought by different codecs. The parameters of RawBoost were set as per \cite{tak2022automatic}.\\
\emph{Silence removal}: Previous works have shown that removing silence frames can lead to drastic degradation to existing deepfake detection models~\cite{liu2023asvspoof}. We therefore trained and evaluated two versions of SLIM, one with silence removed from all samples (train, dev, and eval) and one that kept the silence. Based on the performance achieved on ASV5 {\em prog}, no significant difference was found between the two. Hence, we did not remove silence from the data, when training and evaluating the submitted system.\\
\emph{Zero-padding in batch}: Since samples in the same batch required to be of the same length, we zero-padded shorter samples to align with the length of the longest samples in the batch. To avoid excessive padding, we first sorted all samples in the dataset by length, then batched the samples with similar lengths. This step was applied to training and validation, as both used batch size larger than 1. No zero-padding was applied during inference. However, it should be emphasized that we later found that zero-padding may lead to false acceptance at training time, hence the adopted approach may be sub-optimal. We discuss this issue in Section.~\ref{results:1}.\\
\emph{Training data}: The SSCL pretraining was performed with 3k samples from CommonVoice~\cite{ardila2020common} and 3k samples from RAVDESS~\cite{livingstone2018ryerson}, the entire pretraining took less than \SI{1}{\hour}. Although open condition allows combining data from previous ASVspoof challenges and other sources, we used only ASV5 data for supervised training. \\
\noindent \emph{Hyperparameters}: A summary of all training hyperparameters can be found in Appendix.~\ref{appendix:1}.

\section{Results}
% In this section, we present the performance achieved by SLIM on all employed datasets, including a detailed breakdown of the metrics obtained from ASV5 eval, and the improvements made by different design choices.
\subsection{Model performance}
\label{results:1}
\begin{figure*}
\centering
\includegraphics[clip, trim=3.5cm 9.2cm 3cm 4cm, width=\linewidth]{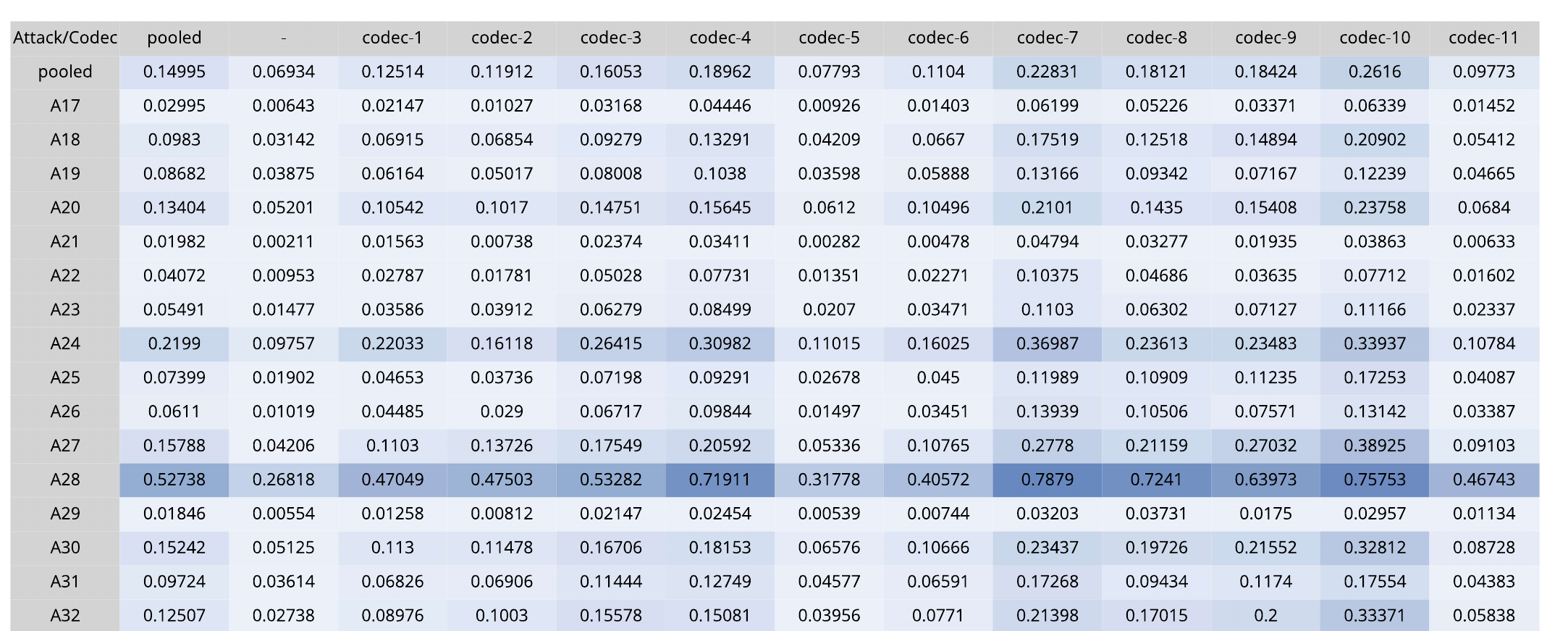}
\caption{Breakdown of system performance (minDCF) on ASV5 {\em eval} dataset.}
\label{fig:minDCF}
\end{figure*}

Our final system achieved an average minDCF of 0.1499 and EER 5.56\% on the ASV5 {\em eval} data. 
Figure.~\ref{fig:minDCF} summarizes the minDCF breakdown for different types of attacks and compression codecs. 
In the clean condition (column 3), our system performs well on 15 of 16 types of 
attacks, obtaining minDCF within 0.1 on all attacks with only one exception (the A28-pretrained YourTTS attack). 

% While A17 and A19 are reported with on average highest minDCF values achieved by track 1 submissions, SLIM was able to accurately detect those two attacks. 
Since all evaluated attacks remain unseen during training, results here suggest that SLIM can generalize well to different generative models, including the ones with adversarial attacks applied (A18, A20, A23, A30--32). This is likely due to our SSCL pretraining on real data. Meanwhile, a significant difference can be seen between the performance achieved on clean data (column 3) and codec-applied (columns 4--12) data. For some codecs, the minDCF values can be 3-5 times higher (e.g., codec-7 and 10). Such 
degradation is expected for our system, considering that we only employed RawBoost for data augmentation. %.  these findings indicate that the pretraining stage may solely enhance the generalizability to unseen generative models while 
Further robustness to unseen codecs can be achieved by employing codec-specific data augmentations. Given that only clean speech data were used for pretraining, improvements may also be achieved by incorporating various data augmentations at the pretraining stage.

\begin{table*}[]
\caption{{\it Comparison of EERs achieved by baseline speech SSL models and SLIM, with ablations of SLIM on different training setups. Overall, consistent improvement is achieved on both Wav2vec and WavLM after SLIM is applied. N/A correspond to scores that remain unknown. ``B\textsubscript{train}'' denotes the batch size during training. Bold values indicate best performance achieved on the given dataset.}}
\vspace{2mm}
\centering
\begin{tabular}{cccccccccc}
\toprule
& SSL Model & DA & Loss & B\textsubscript{train}& ASV5 {\em dev} & ASV2019 & ITW & ASV5 {\em prog} & ASV5 {\em eval} \\
\midrule
\multirow{8}{*}{\makecell{\bf{Cross-model}\\\bf{comparison}}} & WavLM-Base & - & BCE & 8 & 9.6 & 15.2 & 20.0 & 13.4 & NA \\
& WavLM-Base (fft) & - & BCE & 8 & 7.4 & 18.7 & 29.6 & 13.6 & NA\\ 
& Data2vec-Base & - & BCE & 8 & 9.6 & 13.7 & 22.8 & NA & NA \\
& Data2vec-Base (fft) & - & BCE & 8 & 14.6 & 31.1 & 37.6 & NA & NA\\
& Wav2vec-Large & - & BCE & 8 & 7.7 & 15.4 & 22.1 & NA & NA \\
& Wav2vec-Large (fft) & - & BCE & 8 & 18.0 & 25.9  & 35.4 & NA & NA\\
& SLIM (WavLM) & - & BCE & 8 & 5.2 & 11.1 & 25.7 & 7.1 & NA \\
\midrule
\multirow{5}{*}{\makecell{\bf{SLIM}\\\bf{ablation}}}  & SLIM (Wav2vec) & - & BCE & 8 & 7.7 & 12.9 & 19.2 & NA & NA \\
& SLIM (WavLM) & RawBoost & BCE & 8 & \bf 2.9 & 9.5 & \bf 10.8 & 3.6 & NA \\
& SLIM (WavLM) & RawBoost+Noise+RIR & BCE & 8 & 3.3 & 10.4 & 12.4 & NA & NA \\ 
& SLIM (WavLM) & RawBoost & Focal & 8 & 3.8 & 10.7 & 14.5 & 2.7 & NA \\
& SLIM (WavLM) & RawBoost & BCE & 4 & 3.0 & \bf 7.4 & \bf 10.8 & \bf 2.4 & {\bf{5.5}} \\
\bottomrule
\end{tabular}
\label{tab:model_compare}
\end{table*}

% \begin{figure*}
% \centering
% \includegraphics[clip, trim=3.5cm 6.5cm 3cm 7cm, width=\linewidth]{fig/EER.pdf}
% \caption{Breakdown of SLIM's performance (EER) on ASVspoof5 evaluation data.}
% \label{fig:EER}
% \end{figure*}

In Table ~\ref{tab:model_compare}, we show cross-model comparision and ablations on the pre-training stage that led to our choice of SSL encoder and data augmentation. 
%Next, we investigate the impact of the pretraining stage of SLIM by comparing its performance to other baseline SSL models, as presented in Table~\ref{tab:model_compare} - Cross-model comparison.
Since the number of tests on AVspoof5 {\em prog} and {\em eval} sets were limited by the submission quota, we relied on the performance achieved on ASV5 {\em dev}, ASV2019 LA {\em eval}, and ITW datasets to pick the best candidate. All baseline SSL models used the same classifiers  as SLIM, i.e., ASP+MLPs.  The only difference of SLIM was the style-linguistics dependency embeddings learned from Stage 1 pretraining. In general, baseline SSL models (rows 1--6) do not generalize well to unseen attacks, where a marked discrepancy can be seen between the EERs obtained on ASV5 {\em dev} and the out-of-domain datasets. 
SLIM, on the other hand, significantly reduces the generalization gap. For example, with the same base WavLM encoder, baseline model (row 1) achieves EER 13.4\% on ASV5 {\em prog} while SLIM (row 7) achieves 7.1\%. Additionally, though previous works have reported improvements with full finetuning of the SSL encoder~\cite{yi2023audio}, we noticed that full finetuning on ASV5 exacerbates the generalization issue and leads to worse performance than using a frozen backbone.
That said, full finetuning usually requires careful selection of hyperparameters and more compute to train. The results here may not represent the best of full finetuned models due to the time constraint of the challenge.

The ablation part of Table~\ref{tab:model_compare} (lower half) presents other factors that contribute to our choice of candidate. The 7.1\% EER on ASV5 {\em prog} was further decreased to 
3.6\% after RawBoost augmentation was applied. However, no further improvement was seen when combining noise and reverberation augmentations with RawBoost. During training, we noticed 
that the majority of samples can be easily correctly classified after the first epoch, and our model struggled to learn on the remaining hard samples (mostly short and noisy recordings) throughout 
all remaining epochs. This motivated us to substitute BCE loss with Focal loss~\cite{lin2017focal}, as the latter forces the model to focus on hard samples. The loss modification led to another improvement of SLIM which decreases the EER to 2.7\%. However, the focal loss modification was not integrated with our final submitted system (row 12) due to time constraints. Lastly, we found that decreasing batch size from 8 to 4 (before augmentation) also helped with improving the {\em prog} EER by 1.2\%. We conjecture that this may be due to 
the zero-padding performed at the batch level during training, as the spoof and bonafide samples would share the same frames with zero values near the end of recording that may confuse the model. %Nonethless, we left this to be addressed in future works.

Finally, Table~\ref{tab:slim_ablation} summarizes a few factors at pre-processing level that were shown to significantly impact the model performance. As can be seen, longer training samples do not necessarily lead to better performance for ASV5, which partially contradicts the monotonic relationship between audio length and detection performance reported in the previous work~\cite{muller2022does}. One potential cause could be the different distributions of audio lengths for spoof and bonafide in ASV5 {\em train}. Unlike in previous challenges, over 90\% of bonafide samples are found to be more than \SI{10}{\second} long while the majority of spoof samples are shorter than \SI{10}{\second}. This poses a challenge for determining the optimal truncation length of training samples, as taking the entire speech would likely bias the model to believe longer samples are bonafide, while aggressive truncation may lead to significant loss of information. With limited ablation, we found that a sample length of \SI{10}{\second} gave a decent balance. We also experimented with and without silence removal from all partitions and found that removing silence slightly dropped the detection performance (up to 1\% EER difference on ASV5 {\em prog}).

\begin{table}
\caption{{\it Impact of sample length and silence removal on detection performance (in EER).}}
\vspace{2mm}
\centering
\begin{tabular}{cccc|cc}
\toprule
\multirow{2}{*}{Dataset}& \multicolumn{3}{c}{Sample length} & \multicolumn{2}{c}{Silence removal}\\
\cmidrule{2-4} \cmidrule{5-6}
& 8s & 10s & 15s & \cmark & \xmark \\
\midrule
ASV5 {\em dev} & 3.13 & 3.03 & 6.11 & 3.75 & 3.03\\
ASV2019 & 9.9 & 7.4 & 12.8 & 9.5 & 7.4\\
ITW & 14.0 & 10.8 & 16.3 & 13.8 & 10.8\\
\bottomrule
\end{tabular}
\label{tab:slim_ablation}
\end{table}

\subsection{Potential causes of misclassifications}
% We describe the noisy samples and multi-spks distribution in eval and their impact on SLIM decisions here.

\begin{figure}
    \centering
    \includegraphics[width=\linewidth]{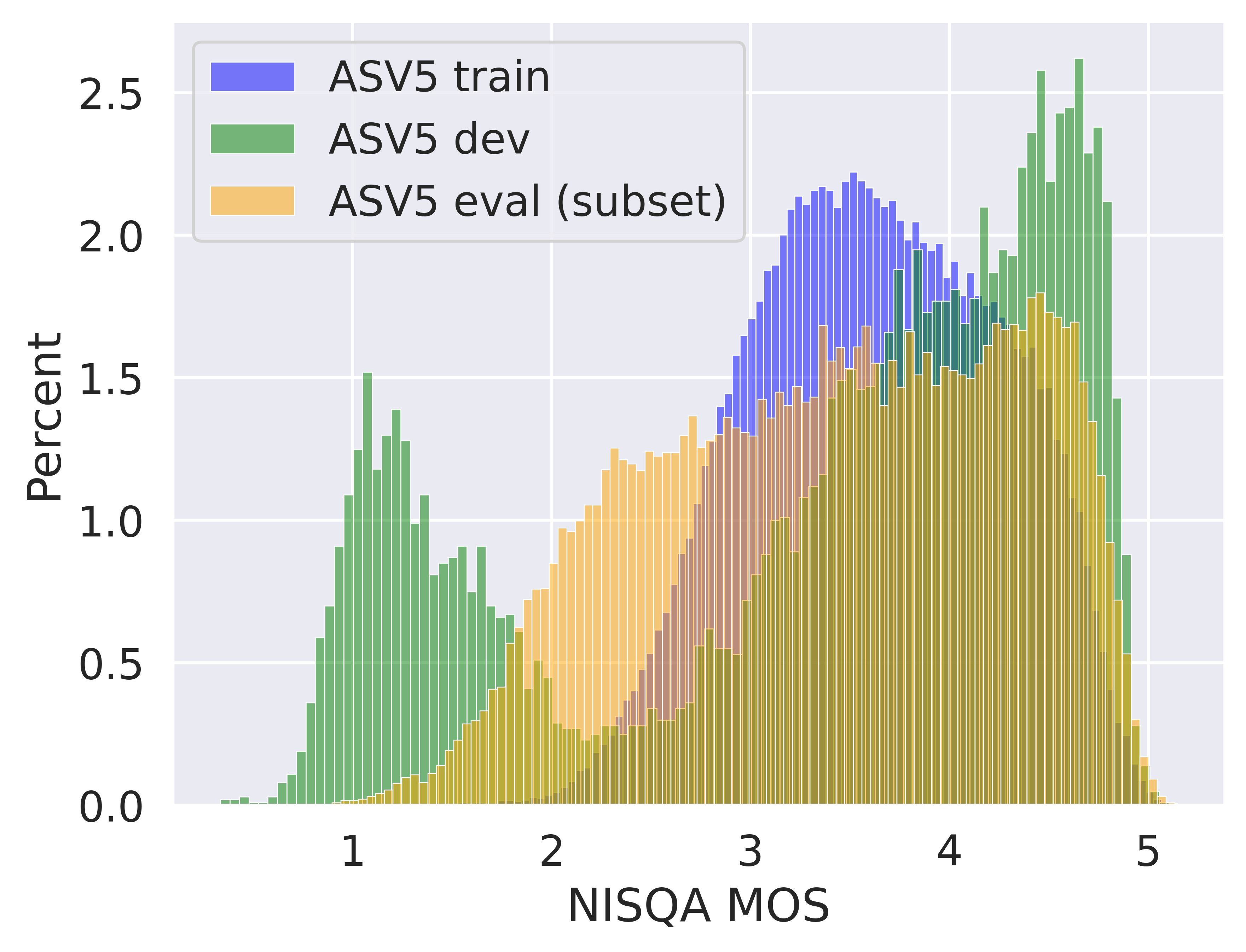}
    \caption{Distribution of NISQA MOS for ASV5 {\em train}, {\em dev}, and 40k samples from {\em eval}.}
    \label{fig:stats}
\end{figure}

% {\color{blue} placeholders for \\(1) Figure: MOS score histograms of the whole dev set and misclassified ones;\\ (2) Figure: spectrograms of the misclassified noisy samples; \\(3) Table: number of speakers in the eval set; \\(4) Figure: shift in the SLIM decision after modifying scores for multi-spk samples.}
While our SLIM model demonstrated good generalizability, it is crucial to investigate the causes of misclassifications. Figure~\ref{fig:stats} depicts the distribution of Mean Opinion Score (MOS) estimated by the NISQA model~\cite{mittag2021nisqa} for ASV5 {\em train}, {\em dev}, and 40k samples from the {\em eval} sets. Training data show significantly higher speech quality than dev and eval sets, where 31.3\% and 33.9\% of dev and eval (subset) samples have NISQA-MOS $\leq$ 3 while only 17.4\% of training data fall in this range. Furthermore, it was found that the misclassified dev samples by SLIM are mostly with NISQA-MOS $\leq$ 3, of which the speech content was nearly unintelligible. Meanwhile, after applying a speaker diarization model on 70k samples randomly selected from the eval set, we noticed that nearly 10\% data may include more than one speaker, representing a conversational or overlapping speech settings. Since the pretraining stage of SLIM was performed within a single speaker scenario, the multi-speaker samples will likely lead to a mismatch detected between style and linguistics, and finally resulting in a misclassification.

\section{Conclusion}
In this paper, we present Reality Defender's submission to the ASVspoof5 challenge. Our submitted system SLIM achieved competitive results on the ASVspoof5 eval set as well as two out-of-domain deepfake datasets. Our findings suggest that the self-supervised contrastive learning stage of SLIM can effectively improve the generalizability to unseen attacks. Further research is needed to  improve the performance in a multi-speaker setting, and for robustness to specific compression codecs.

\section{Acknowledgement}
%Authors would like to acknowledge the organizers of ASVspoof challenges as well as the data holders of the datasets employed in this study.
A part of this research was supported by Compute Canada \cite{baldwin2012compute}. The authors would like to thank Brahmi Dwivedi and Franscesco Nespoli for their data investigation and early brainstorming on the ASV5 challenge.

\section{Appendix}
\subsection{Hyperparameters}
\label{appendix:1}
Table~\ref{tab:hparam} describes the hyperparameter settings and the architecture details of SLIM for Stage 1 and Stage 2 training. Both training stages were performed using SpeechBrain \textit{v1.0.0}. Experiments were conducted on the Compute Canada cluster with four NVIDIA V100 GPUs (32GB RAM) for Stage 1 and one A100 (40GB RAM) for Stage 2.

\begin{table}[hbpt]
\caption{Hyperparameters and architecture details of SLIM.}
\label{tab:hparam}
\centering
%\begin{tabularx}{0.58\linewidth}{cc}
\begin{tabular}{cc}
\toprule
{\bf{Parameter}} & {\bf{Setting}} \\
\midrule \midrule
\multicolumn{2}{c}{\bf{Stage 1 Optimization}} \\
Batch size & 16 \\
Epochs & 50 \\
GPUs & 4 V100 \\
Audio length & 10s \\
Optimizer & AdamW \\
LRscheduler & Linear \\
Starting LR & .005 \\
End LR & .0001 \\
Early-stop patience & 3 epochs \\
$\lambda$ & .007 \\
Training time & 1h\\
\midrule
\multicolumn{2}{c}{\bf{SSL frontend}}\\
Style encoder & WavLM Base (0-7 layers) \\
Linguistic encoder & WavLM Base (8-11 layers)\\
\midrule
\multicolumn{2}{c}{\bf{Projector}}\\
Bottleneck layers & 1 \\
BN dropout & 0.1 \\
FC dropout & 0.1 \\
Compression output dim & 256 \\
\midrule
\multicolumn{2}{c}{\bf{Stage 2 Optimization}} \\
Batch size & 4 (8 after DA) \\
Epochs & 5 \\
% GPUs & 1 A100 \\
Audio length & 10s \\
BCE class weight & 10:1 \\
Optimizer & AdamW \\
LRscheduler & Linear \\
Starting LR & .001 \\
End LR & .0001 \\
Early-stop patience & 3 epochs \\
Training time & 14h (1 A100 GPU) \\
\midrule
\multicolumn{2}{c}{\bf{Classifier}} \\
FC dropout & 0.25 \\
\midrule
\multicolumn{2}{c}{\bf{Stage 2}} \\
Augmentations & RawBoost \\
Concat with original & True \\
\bottomrule
\end{tabular}
\end{table}

\subsection{Projector and classifier architecture}
\label{appendix:2}
\begin{figure}[hbpt]
\centering
\includegraphics[width=0.9\linewidth]{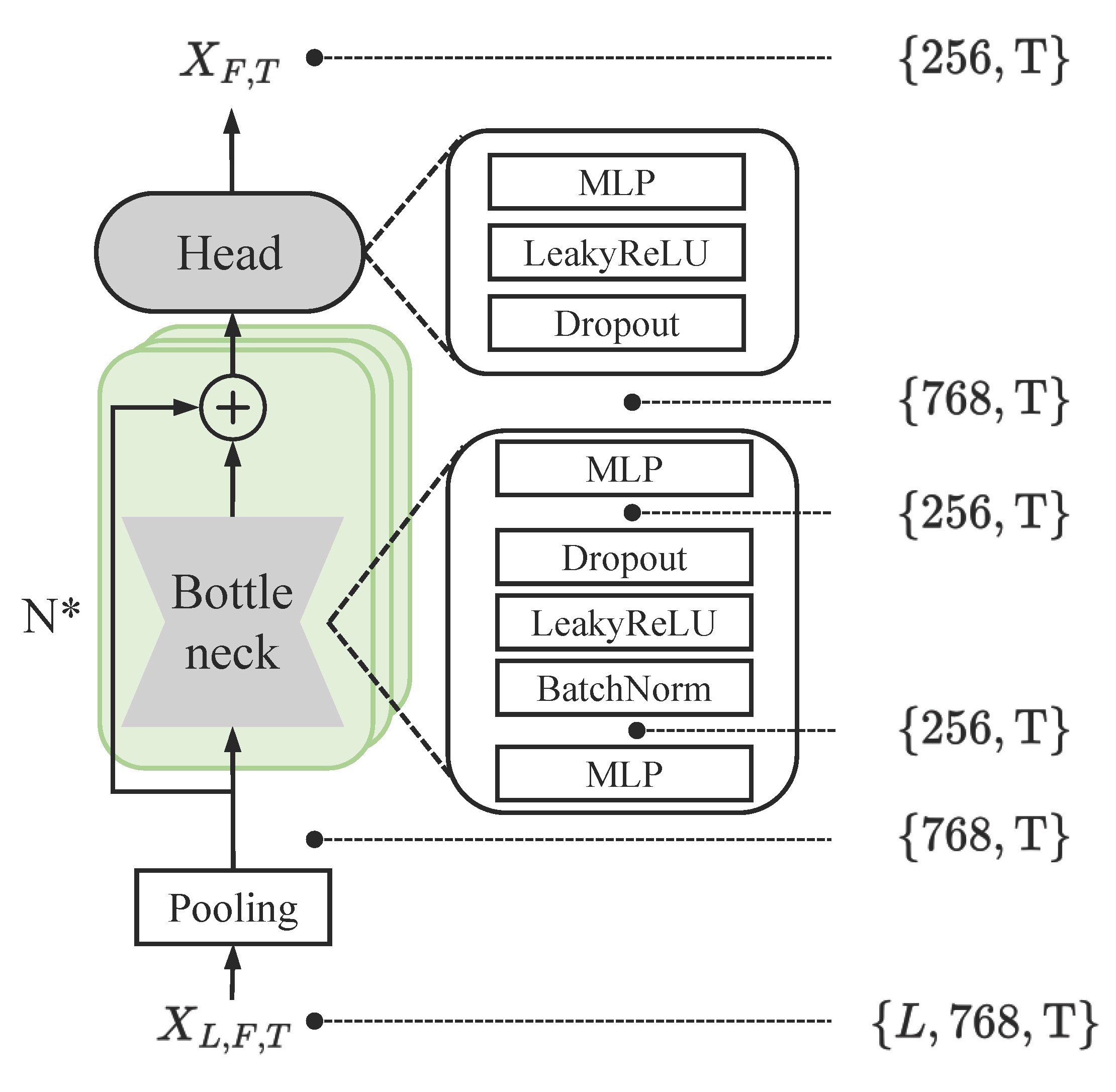}
\caption{Architecture of the projector network with input and output dimensions. Input $\mathbf{X_{L,F,T}}$ represents the original subspace representation encoded by the SSL frontend, where $L$ denotes the transformer layer index, $F$ denotes the feature size, and $T$ denotes the number of time steps.}
\label{fig:compress}
\end{figure}

Figure~\ref{fig:compress} shows the architecture of the projector network. The input is first passed through a pooling layer to obtain an average of different layer outputs. Bottleneck layers are designed to remove the redundant information that is not shared across style and linguistics aspects. The bottleneck layer first maps the feature dimension from 768-dim to 256-dim, then recovers it back to 768-dim. In practice, we found using only one bottleneck layer is enough to obtain meaningful compressed representations. A projection head is applied at the end to reduce the final output dimension to 256. The 256-dim embeddings then pass through a fully-connected layer, a dropout layer, and a final projection layer that yields a single output score.

\bibliographystyle{IEEEbib}
\bibliography{ASVspoof_BibEntries}

\begin{thebibliography}{10}

\bibitem{almutairi2022review}
Zaynab Almutairi and Hebah Elgibreen,
\newblock ``A review of modern audio deepfake detection methods: Challenges and future directions,''
\newblock {\em Algorithms}, vol. 15, no. 5, pp. 155, 2022.

\bibitem{masood2023deepfakes}
Momina Masood, Mariam Nawaz, Khalid~Mahmood Malik, Ali Javed, Aun Irtaza, and Hafiz Malik,
\newblock ``Deepfakes generation and detection: State-of-the-art, open challenges, countermeasures, and way forward,''
\newblock {\em Applied intelligence}, vol. 53, no. 4, pp. 3974--4026, 2023.

\bibitem{news1}
Kate Knibbs,
\newblock ``Researchers say the deepfake biden robocall was likely made with tools from ai startup elevenlabs,'' \url{https://www.wired.com/story/biden-robocall-deepfake-elevenlabs/}, 2024,
\newblock Accessed: 2024-04-30.

\bibitem{news2}
Joseph Cox,
\newblock ``How i broke into a bank account with an ai-generated voice,'' \url{https://www.vice.com/en/article/dy7axa/how-i-broke-into-a-bank-account-with-an-ai-generated-voice}, 2024,
\newblock Accessed: 2024-07-26.

\bibitem{news3}
EDRM Electronic Discovery~Reference Model,
\newblock ``Ai threatens courts with fake evidence, uw prof says,'' \url{https://www.jdsupra.com/legalnews/ai-threatens-courts-with-fake-evidence-7371356/}, 2024,
\newblock Accessed: 2024-07-26.

\bibitem{todisco2019asvspoof}
Massimiliano Todisco, Xin Wang, Ville Vestman, Md~Sahidullah, H{\'e}ctor Delgado, Andreas Nautsch, Junichi Yamagishi, Nicholas Evans, Tomi Kinnunen, and Kong~Aik Lee,
\newblock ``{ASVspoof 2019}: Future horizons in spoofed and fake audio detection,''
\newblock {\em arXiv preprint arXiv:1904.05441}, 2019.

\bibitem{liu2023asvspoof}
Xuechen Liu, Xin Wang, Md~Sahidullah, Jose Patino, H{\'e}ctor Delgado, Tomi Kinnunen, Massimiliano Todisco, Junichi Yamagishi, Nicholas Evans, Andreas Nautsch, et~al.,
\newblock ``{ASVspoof 2021}: Towards spoofed and deepfake speech detection in the wild,''
\newblock {\em IEEE/ACM Transactions on Audio, Speech, and Language Processing}, 2023.

\bibitem{9747768}
Juan~M. Martín-Doñas and Aitor Álvarez,
\newblock ``{The Vicomtech Audio Deepfake Detection System Based on Wav2vec2 for the 2022 ADD Challenge},''
\newblock in {\em IEEE International Conference on Acoustics, Speech and Signal Processing (ICASSP)}, 2022, pp. 9241--9245.

\bibitem{tak2022automatic}
Hemlata Tak, Massimiliano Todisco, Xin Wang, Jee-weon Jung, Junichi Yamagishi, and Nicholas Evans,
\newblock ``Automatic speaker verification spoofing and deepfake detection using wav2vec 2.0 and data augmentation,''
\newblock in {\em The Speaker and Language Recognition Workshop (Odyssey 2022)}. ISCA, 2022.

\bibitem{jung2022aasist}
Jee-weon Jung, Hee-Soo Heo, Hemlata Tak, Hye-jin Shim, Joon~Son Chung, Bong-Jin Lee, Ha-Jin Yu, and Nicholas Evans,
\newblock ``Aasist: Audio anti-spoofing using integrated spectro-temporal graph attention networks,''
\newblock in {\em IEEE international conference on acoustics, speech and signal processing (ICASSP)}. IEEE, 2022, pp. 6367--6371.

\bibitem{kang2024experimental}
Taein Kang, Soyul Han, Sunmook Choi, Jaejin Seo, Sanghyeok Chung, Seungeun Lee, Seungsang Oh, and Il-Youp Kwak,
\newblock ``Experimental study: Enhancing voice spoofing detection models with wav2vec 2.0,''
\newblock {\em arXiv preprint arXiv:2402.17127}, 2024.

\bibitem{wang2023spoofed}
Xin Wang and Junichi Yamagishi,
\newblock ``Spoofed training data for speech spoofing countermeasure can be efficiently created using neural vocoders,''
\newblock in {\em IEEE International Conference on Acoustics, Speech and Signal Processing (ICASSP)}. IEEE, 2023, pp. 1--5.

\bibitem{wang2024can}
Xin Wang and Junichi Yamagishi,
\newblock ``Can large-scale vocoded spoofed data improve speech spoofing countermeasure with a self-supervised front end?,''
\newblock in {\em IEEE International Conference on Acoustics, Speech and Signal Processing (ICASSP)}. IEEE, 2024, pp. 10311--10315.

\bibitem{tak2022rawboost}
Hemlata Tak, Madhu Kamble, Jose Patino, Massimiliano Todisco, and Nicholas Evans,
\newblock ``Rawboost: A raw data boosting and augmentation method applied to automatic speaker verification anti-spoofing,''
\newblock in {\em IEEE International Conference on Acoustics, Speech and Signal Processing (ICASSP)}. IEEE, 2022, pp. 6382--6386.

\bibitem{yang2024robust}
Yujie Yang, Haochen Qin, Hang Zhou, Chengcheng Wang, Tianyu Guo, Kai Han, and Yunhe Wang,
\newblock ``A robust audio deepfake detection system via multi-view feature,''
\newblock in {\em IEEE International Conference on Acoustics, Speech and Signal Processing (ICASSP)}. IEEE, 2024, pp. 13131--13135.

\bibitem{Wang2024_ASVspoof5}
X.~Wang et~al.,
\newblock ``{ASVspoof 5}: Crowdsourced data, deepfakes and adversarial attacks at scale,''
\newblock in {\em ASVspoof 2024 workshop (submitted)}, 2024.

\bibitem{zhu2024slim}
Yi~Zhu, Surya Koppisetti, Trang Tran, and Gaurav Bharaj,
\newblock ``{SLIM}: Style-linguistics mismatch model for generalized audio deepfake detection,''
\newblock {\em arxiv preprint arXiv:2407.18517}, 2024.

\bibitem{muller2022does}
Nicolas M{\"u}ller, Pavel Czempin, Franziska Diekmann, Adam Froghyar, and Konstantin B{\"o}ttinger,
\newblock ``Does audio deepfake detection generalize?,''
\newblock {\em INTERSPEECH}, 2022.

\bibitem{schuller2013paralinguistics}
Bj{\"o}rn Schuller, Stefan Steidl, Anton Batliner, Felix Burkhardt, Laurence Devillers, Christian M{\"u}Ller, and Shrikanth Narayanan,
\newblock ``Paralinguistics in speech and language—state-of-the-art and the challenge,''
\newblock {\em Computer Speech \& Language}, vol. 27, no. 1, pp. 4--39, 2013.

\bibitem{kretzschmar2009linguistics}
William~A Kretzschmar,
\newblock {\em The linguistics of speech},
\newblock Cambridge University Press, 2009.

\bibitem{lindquist2015role}
Kristen~A Lindquist, Jennifer~K MacCormack, and Holly Shablack,
\newblock ``The role of language in emotion: Predictions from psychological constructionism,''
\newblock {\em Frontiers in psychology}, vol. 6, pp. 121301, 2015.

\bibitem{cutler1997prosody}
Anne Cutler, Delphine Dahan, and Wilma Van~Donselaar,
\newblock ``Prosody in the comprehension of spoken language: A literature review,''
\newblock {\em Language and speech}, vol. 40, no. 2, pp. 141--201, 1997.

\bibitem{pereira2019age}
Natalie Pereira, Ana Paula~Bresolin Gon{\c{c}}alves, Mariana Goulart, Marina~Amarante Tarrasconi, Renata Kochhann, and Rochele~Paz Fonseca,
\newblock ``Age-related differences in conversational discourse abilities a comparative study,''
\newblock {\em Dementia \& Neuropsychologia}, vol. 13, pp. 53--71, 2019.

\bibitem{baevski2020wav2vec}
Alexei Baevski, Yuhao Zhou, Abdelrahman Mohamed, and Michael Auli,
\newblock ``Wav2vec 2.0: A framework for self-supervised learning of speech representations,''
\newblock {\em Advances in neural information processing systems}, vol. 33, pp. 12449--12460, 2020.

\bibitem{chen2022wavlm}
Sanyuan Chen, Chengyi Wang, Zhengyang Chen, Yu~Wu, Shujie Liu, Zhuo Chen, Jinyu Li, Naoyuki Kanda, Takuya Yoshioka, Xiong Xiao, et~al.,
\newblock ``{WavLM}: Large-scale self-supervised pre-training for full stack speech processing,''
\newblock {\em IEEE Journal of Selected Topics in Signal Processing}, vol. 16, no. 6, pp. 1505--1518, 2022.

\bibitem{baevski2022data2vec}
Alexei Baevski, Wei-Ning Hsu, Qiantong Xu, Arun Babu, Jiatao Gu, and Michael Auli,
\newblock ``Data2vec: A general framework for self-supervised learning in speech, vision and language,''
\newblock in {\em International Conference on Machine Learning}. PMLR, 2022, pp. 1298--1312.

\bibitem{cho2023evidence}
Cheol~Jun Cho, Peter Wu, Abdelrahman Mohamed, and Gopala~K Anumanchipalli,
\newblock ``Evidence of vocal tract articulation in self-supervised learning of speech,''
\newblock in {\em IEEE International Conference on Acoustics, Speech and Signal Processing (ICASSP)}. IEEE, 2023, pp. 1--5.

\bibitem{pasad2023comparative}
Ankita Pasad, Bowen Shi, and Karen Livescu,
\newblock ``Comparative layer-wise analysis of self-supervised speech models,''
\newblock in {\em IEEE International Conference on Acoustics, Speech and Signal Processing (ICASSP)}. IEEE, 2023, pp. 1--5.

\bibitem{saliba2024layer}
Alexandra Saliba, Yuanchao Li, Ramon Sanabria, and Catherine Lai,
\newblock ``{Layer-Wise Analysis of Self-Supervised Acoustic Word Embeddings: A Study on Speech Emotion Recognition},''
\newblock {\em arXiv preprint arXiv:2402.02617}, 2024.

\bibitem{ardila2020common}
Rosana Ardila, Megan Branson, Kelly Davis, Michael Kohler, Josh Meyer, Michael Henretty, Reuben Morais, Lindsay Saunders, Francis Tyers, and Gregor Weber,
\newblock ``{Common Voice}: A massively-multilingual speech corpus,''
\newblock in {\em Proceedings of the Twelfth Language Resources and Evaluation Conference}, 2020, pp. 4218--4222.

\bibitem{livingstone2018ryerson}
Steven~R Livingstone and Frank~A Russo,
\newblock ``The ryerson audio-visual database of emotional speech and song {(RAVDESS)}: A dynamic, multimodal set of facial and vocal expressions in north american english,''
\newblock {\em PloS one}, vol. 13, no. 5, pp. e0196391, 2018.

\bibitem{yang2022torchaudio}
Yao-Yuan Yang, Moto Hira, Zhaoheng Ni, Artyom Astafurov, Caroline Chen, Christian Puhrsch, David Pollack, Dmitriy Genzel, Donny Greenberg, Edward~Z Yang, et~al.,
\newblock ``Torchaudio: Building blocks for audio and speech processing,''
\newblock in {\em IEEE International Conference on Acoustics, Speech and Signal Processing (ICASSP)}. IEEE, 2022, pp. 6982--6986.

\bibitem{yi2023audio}
Jiangyan Yi, Chenglong Wang, Jianhua Tao, Xiaohui Zhang, Chu~Yuan Zhang, and Yan Zhao,
\newblock ``Audio deepfake detection: A survey,''
\newblock {\em arXiv preprint arXiv:2308.14970}, 2023.

\bibitem{lin2017focal}
Tsung-Yi Lin, Priya Goyal, Ross Girshick, Kaiming He, and Piotr Doll{\'a}r,
\newblock ``Focal loss for dense object detection,''
\newblock in {\em Proceedings of the IEEE international conference on computer vision}, 2017, pp. 2980--2988.

\bibitem{mittag2021nisqa}
Gabriel Mittag, Babak Naderi, Assmaa Chehadi, and Sebastian M{\"o}ller,
\newblock ``{NISQA}: A deep cnn-self-attention model for multidimensional speech quality prediction with crowdsourced datasets,''
\newblock {\em arXiv preprint arXiv:2104.09494}, 2021.

\bibitem{baldwin2012compute}
Susan Baldwin,
\newblock ``{Compute Canada}: advancing computational research,''
\newblock in {\em Journal of Physics: Conference Series}. IOP Publishing, 2012, vol. 341, p. 012001.

\end{thebibliography}

% This could be also done as follows:
%
%\begin{thebibliography}{10}
%\bibitem[1]{aluisio2001learn}Sandra M. Alu\'{i}sio, Iris Barcelos, Jandir Sampaio, and Osvaldo
%N. Oliveira Jr, ``How to learn the many unwritten
%``rules of the game'' of the academic discourse: a hybrid
%approach based on critiques and cases to support scientific
%writing,'' in Proceedings of the IEEE International Conference
%on Advanced Learning Technologies, Madison, USA,
%August 2001, pp. 257–260.
%\bibitem[2]{swales1987writing} John Swales and Hazem Najjar, ``The writing of research
%article introductions,'' Written communication, vol. 4, no.
%2, pp. 175–191, 1987.
%\bibitem[3]{day2012write} Robert Day and Barbara Pastel, How to write and publish
%a scientific paper, Cambridge University Press, 2012.
%\bibitem[4]{teufel2000} Simone Teufel, Argumentative zoning: information extraction
%from scientific text, Ph.D. thesis, University of Edinburgh,
%2000.
%\bibitem[5]{berkenkotter1989social} Carol Berkenkotter, Thomas N. Huckin, and John Ackerman,
%``Social context and socially constructed texts: The
%initiation of a graduate student into a writing research community.
%technical report no. 33.,'' Tech. Rep., Center for
%the Study of Writing, University of California Berkeley \&
%Carnegie Mellon University, 1989.
%\end{thebibliography}

\end{document}